\documentstyle[11pt]{article}
\newcommand{\bxi}{\mbox{\boldmath $\xi$}}
\newcommand{\bsigma}{\mbox{\boldmath $\sigma$}}
\def \beq{\begin{equation}}
\def \eeq{\end{equation}}
\def\be{ \begin{displaymath} }
\def\ee{ \end{displaymath} }
\def\ben{ \begin{equation} }
\def\een{ \end{equation} }
\def\bea{ \begin{eqnarray} }
\def\eea{ \end{eqnarray} }

\def\ko{ {\rm K}^ 0 }
\def\kob{ \overline{\rm K}{}^0 }
\def\kak{ {\rm K}^ 0 - \overline{\rm K}{}^0 }
\def\kl{ {\rm K_{L}} }
\def\ks{ {\rm K_{S}} }

\def\kk{ | {\rm K^ 0} \rangle }
\def\kbk{ | {\rm \overline{K}{}^0} \rangle }
\def\koo{ | {\rm K^ 0}(0) \rangle }
\def\bkoo{ | {\rm \overline{K}{}^0}(0) \rangle }
\def\kkl{|K_L^0>}

\def\kks{|K_S^0>}

\def\kku{|K_1^0>}

\def\kkd{|K_2^0>}

\def\kst{|K_S(t)>}
\def\kso{|K_S(0)>}
\def\klt{|K_L(t)>}
\def\klo{|K_L(0)>}
\def\kot{|K^0(t)>}
\def\bkot{|{\overline K}^0(t)>}

\def\eas{e^{-i \lambda_S t}}
\def\eal{e^{-i \lambda_L t}}
\def\fpmt{f_{\pm}(t)}
\def\fpt{f_{+}(t)}
\def\fmt{f_{-}(t)}

\def \cp {\par \noindent }
 
\def \kok {K^0- \overline {K^0} }

\def \ee {${\epsilon ' \over \epsilon} $ }

\def \ko {K^0 }
\def \kbo {\overline{K^0} }

\def\vec#1{{\bf #1}}

\def\ext{_{\rm ext}}

\def\Iscr{{\cal I}}

\def\Rscr{{\cal R}}
\def\Hscr{{\cal H}}

\def\au{{\alpha_1}}
\def\ad{{\alpha_2}}
\def\Re{\rm Re}

\newcommand{\Ebo}{\mbox{{\boldmath $E$}}}
\newcommand{\FEbo}{\mbox{{\boldmath $\widetilde E$}}}

\newcommand{\epsx}{\mbox{{\boldmath $\epsilon_1$}}}
\newcommand{\epsy}{\mbox{{\boldmath $\epsilon_2$}}} 
\newcommand{\epsl}{\mbox{{\boldmath $\epsilon_{-}$}}}
\newcommand{\epsr}{\mbox{{\boldmath $\epsilon_{+}$}}} 

\def\ket#1{\vert #1 \rangle }
\def\modu#1{\vert {\vec #1} \vert }

\def\Ds {D \kern-2.2ex /} 
\def\dw {W \kern-2.2ex /} 
\def\dz {Z \kern-2.2ex /}
\def\dk {k \kern-1.2ex /}
\newcommand{\lsim}{\mathrel{\raise.5ex\hbox{$<$}\kern-.75em\raise
					     -.5ex\hbox{$\sim$}}}
\newcommand{\gsim}{\mathrel{\raise.5ex\hbox{$>$}\kern-.75em\raise
					     -.5ex\hbox{$\sim$}}}
\newskip\humongous \humongous=0pt plus 1000pt minus 1000pt
\def\caja{\mathsurround=0pt}
\def\eqalign#1{\,\vcenter{\openup1\jot \caja
	\ialign{\strut \hfil$\displaystyle{##}$&$
	\displaystyle{{}##}$\hfil\crcr#1\crcr}}\,}
\newif\ifdtup

\catcode`\@=11
\def\section{\@startsection {section}{1}{0pt}{-3.5ex plus -1ex minus
 -.2ex}{2.3ex plus .2ex}{\raggedright\large\bf}}
\catcode`\@=12
\catcode`\@=11
\def\eqnarray{\stepcounter{equation}\let\@currentlabel=\theequation
\global\@eqnswtrue
\global\@eqcnt\z@\tabskip\@centering\let\\=\@eqncr
\gdef\@@fix{}\def\eqno##1{\gdef\@@fix{##1}}%
$$\halign to \displaywidth\bgroup\@eqnsel\hskip\@centering
  $\displaystyle\tabskip\z@{##}$&\global\@eqcnt\@ne
  \hskip 2\arraycolsep \hfil${##}$\hfil
  &\global\@eqcnt\tw@ \hskip 2\arraycolsep $\displaystyle\tabskip\z@{##}$\hfil
   \tabskip\@centering&\llap{##}\tabskip\z@\cr}

\def\@@eqncr{\let\@tempa\relax
    \ifcase\@eqcnt \def\@tempa{& & &}\or \def\@tempa{& &}
      \else \def\@tempa{&}\fi
     \@tempa \if@eqnsw\@eqnnum\stepcounter{equation}\else\@@fix\gdef\@@fix{}\fi
     \global\@eqnswtrue\global\@eqcnt\z@\cr}
\begin{document}
\renewcommand{\thetable}{\Roman{table}}

\setlength{\baselineskip}{0.8 cm}

\rightline{UNIBAS--MATH 11/96}
\vspace{2.5cm}

\begin{center}
{\Large \bf THE OPTICAL ANALOGUE OF CP--VIOLATION}
\end{center}
\vspace{1cm}

\begin{center}
{\large D. Cocolicchio$^{(1,2)}$, L. Telesca$^{(3)}$ and M. Viggiano$^{(1)}$ }
\end{center}
\vspace{0.3cm}
\begin{center}
$^{(1)}$
{\it
Dipartimento di Matematica, Univ. Basilicata, Potenza, Italy\\
Via N. Sauro 85, 85100 Potenza, Italy}
\end{center}

\begin{center}
$^{(2)}$
{\it
Istituto Nazionale di Fisica Nucleare, Sezione di Milano, Italy\\
Via G. Celoria 16, 20133 Milano, Italy}
\end{center}
\begin{center}
$^{(3)}$
{\it
Consiglio Nazionale delle Ricerche, Istituto di Metodologie 
Avanzate\\
C/da S. Loya, 85050 Tito, Potenza, Italy}
\end{center}
\vspace{0.5cm}
\begin{center}
P.A.C.S. number(s): 
11.30.Er,~~     
13.20.Eb,~~     
13.25.+m,~~     
14.40.Aq        
\end{center}
\vspace{1.5cm}
\begin{abstract}
\noindent
The peculiar features of the mixing in the neutral 
pseudoscalar mesons $\kok$ can be introduced by the analogy to the 
optical polarization.
The time-reversed not-invariant processes and the related 
phenomenon of $CP$-nonconservation can be then joined to the
dissipative effects which yield a not vanishing imaginary part in the 
relevant propagation of electromagnetic radiation in a   medium.
Thus, the propagation of the two transverse
polarization states can reproduce the
peculiar asymmetries which are so common in the realm
of high energy particle physics.
\end{abstract}
\normalsize

\vfill~\vfill~
\thispagestyle{empty}
\newpage
\baselineskip=12pt
\section{Introduction}

\bigskip
\noindent
Mechanical models that provide a viable description of the
non-unitary time evolution can be related to the problem of two coupled 
degenerate oscillators with dissipation~\cite{LT}.
The physical intuition underlying any proposed model stems back from the fact
that any time-reversed non invariant process can yield a non vanishing
imaginary part for the relevant Hamiltonian.
Although, the Hamiltonian of a complete sensible quantum system is expected
to be a Hermitian operator, under suitable conditions we may recover the time
evolution of a subsystem according to an effective non hermitian Hamiltonian like in the
case of metastable states \cite{wiger-weis}. A celebrated example where this
description has proved extremely useful is the two-states kaon complex
\cite{LOY}.
In this case, the single pole approximation of the Weisskopf-Wigner method
\cite{wiger-weis} can be applied to derive 
the eigenstates  $K_S$ and $K_L$
of a $2 \times 2$ effective Hamiltonian 
${\cal H}$~\cite{Buras}. 
Although such puzzle system requires the formalism of
density matrix, and the notion of the rest frame for an 
oscillating unstable composite system appears difficult to be implemented, nevertheless
there has been always a steady activity towards understanding these processes at
the level of wave-function~\cite{Buras}.
The time evolution of the flavour states $\ko$ and $\kbo$ can be
written as
\beq
i\hbar {d\over {dt}}
\left( \matrix{ \vert \ko (t)\rangle \cr 
	  \vert \kbo (t) \rangle \cr}\right) 
={\cal H}
\left( \matrix{ \vert \ko (t)\rangle \cr 
	  \vert \kbo (t) \rangle \cr}\right) 
= \left(
\matrix {H_{11} & H_{12} \cr
H_{21} & H_{22} \cr}
\right)
\left( \matrix{ \vert \ko (t)\rangle \cr 
	  \vert \kbo (t) \rangle \cr}\right) 
\eeq
\cp
where the effective $2\times 2$ matrix Hamiltonian ${\cal H}$ 
is linear but not necessarily Hermitian and 
it can be 
uniquely rewritten in terms of two Hermitian matrices
$M$ and $\Gamma$: ${\cal H} = M - i \Gamma /2$.
M is called the mass matrix and $\Gamma$ the decay matrix.
Their explicit expressions can be derived by the weak scattering theory
responsible of the decay~\cite{Buras}.
The dynamical behaviour of this system can be
described in the Schrodinger picture by means of a time evolution
operator $O$:
\beq \label{eqn:eIt}
\left( \matrix{ \vert \ko (t)\rangle \cr 
	  \vert \kbo (t) \rangle \cr}\right) 
=  O(t) 
\left( \matrix{ \vert \ko \rangle \cr 
	  \vert \kbo \rangle \cr}\right) \quad ,
\eeq
which can be written by means of the well known exponential solution
\beq\label{eqn:eItb}
O(t) = \left\{ \exp \left[ -{i\over {\hbar} }
\Hscr t\right] \right\}  ,
\eeq
whose matrix elements are given by
\beq \label{eqn:Uij}
U_{ij} = \langle K_i \vert 
\exp \left[ -{i\over {\hbar} } \Hscr t\right] 
\vert K_j \rangle
\eeq
where as usual the latin indices denote $\ko$, $\kbo$.
In order to evaluate this matrix exponential form, we can use the Sylvester's
formula of the matrix spectral decomposition for a function $f(A)$ of a
generic $n\times n$ square matrix A. 
\noindent
It is easy to express this Sylvester's formula for the special
case of $2\times 2$ matrix with two distinct complex eigenvalues
$\alpha_1$ and $\alpha_2$
\beq \label{eqn:eIq}
f(A) = - {{\ad\au}\over{\ad - \au}}
			   \left[ {{f(\ad)}\over\ad} - {{f(\au)}\over\au} \right] \Iscr +
			  {{1}\over{\ad-\au}} \left[ f(\ad) - f(\au) \right] A 
\eeq
\noindent
which specializes in the limit $\ad\rightarrow\au$ in
\beq \label{eqn:eIqb}
f(A) \rightarrow - \au^2 {d\over{d\au}} 
\left[ {{f(\au)}\over \au}\right]  \Iscr + {{d f(\au)}\over{d\au}}  A
= f(\au) \Iscr + f^\prime(\au) (A-\au \Iscr) 
\eeq

\noindent
This formula becomes rewarding in connection with the expansion of the
exponential representation of the evolution operator.
In fact, in this case, we get

\beq \label{eqn:eIc}
O(t) =  \exp \left( - {i\over {\hbar} } \Hscr t\right) =
{{1}\over{\ad-\au}} \left\{\left[ e^{- {i\over {\hbar}} \ad t } -
e^{- {i\over {\hbar}} \au t } \right] \Hscr + 
\left[ \ad e^{- {i\over {\hbar}} \au t } - \au
e^{- {i\over {\hbar}} \ad t } \right] \Iscr \right\} \, .  
\eeq
It is convenient to represent the matrix $\Hscr$ in terms of the Pauli spin
matrices
\beq \label{eqn:ehu}
{\cal H} = h_\mu\sigma_\mu = h_0 \Iscr + {\vec h}\cdot {
\bsigma}
\eeq
where we let
\beq \label{eqn:ehht}
\eqalign{
h_0= &{1\over 2} (H_{11} + H_{22}) \cr
h_1= &{1\over 2} (H_{12} + H_{21}^*)  \cr
h_2= &{i\over 2} (H_{12} - H_{21}^*) \cr
h_3= &{1\over 2} (H_{11} - H_{22}) \cr}
\eeq
\noindent
being tr${\cal H}=\au+\ad=2 h_0$, $\det {\cal H} = \au \ad = (h_0)^2 - 
\modu{h}^2$
and then $\ad - \au =2 \modu{h}$. It is easy then proven that $H$ is a
normal matrix with a complete set of orthogonal eigenvectors if and only if 
${\vec h} \times {\vec h^* }= 0$.
Rewriting the solution in terms of the $\vec h$ vector, we have
\beq
\eqalign{
O(t) =&\exp \left[ -{i\over {\hbar} }\Hscr t\right] =
 \exp \left(-{i\over {\hbar} } h_0 t\right)  
 \exp \left( -{i\over {\hbar} } {\vec h}\cdot {\bsigma}  t\right) 
=\cr
=&{1\over 2} \left[ \left( 1+{\hat {\vec h}}\cdot{\bsigma}\right) 
e^{- {i\over {\hbar}} \ad t } +
\left( 1-{\hat {\vec h}}\cdot{\bsigma}\right) 
e^{- {i\over {\hbar}} \au t}\right] }
\eeq
where ${\hat {\vec h}}={\vec h}/\modu{h}$.
Except for the overall phase $\exp \left(- i h_0 t\right) $,
this is just a rotation
$ \Rscr [\omega t] = 
\exp\left[  -i ({\vec \omega} \cdot {\vec \sigma} t/2) \right] $ of the
initial state $\ket{\psi(0)}$:
\beq \label{eqn:ehq}
\ket{\psi(t)} = e^{- i h_0 t} \Rscr [\omega t] \ket{\psi(0)} \; .
\eeq
\noindent 
In particle physics, $K^0$ and $\kob$ are expected to be distinct 
particles from the point of view of the strong interactions, they 
could transform into each other through the action of the weak 
interactions. In fact, the system $\kok$  results degenerate due to
a coupling to common final states (direct $CP$-violation)
($K^0\leftrightarrow\pi\pi\leftrightarrow {\kob} $) or by means of a 
mixing ($K^0 \leftrightarrow \kob $) (indirect $CP$ 
violation). Even if this is an effect of second order in the weak interactions, 
the transition from the $K^0$ to $\kob$
becomes important just because this interference in evaluating 
decay widths.
The mass eigenstates are linear combinations
of the flavour states $\kk$  and $\kbk$ namely $\kks$ and $\kkl$,
which have definite masses and lifetimes.
A convenient way to discuss the $\kak$ mixing problem is then to express $\ks$
and $\kl$ as eigenstates of the 2$\times$2 effective Hamiltonian.
It can be shown that if CPT holds then the restriction ${\cal H}_{11} 
= {\cal H}_{22}$ and hence $M_{11} = M_{22}, ~ \Gamma_{11} =
\Gamma_{22}$ must be adopted. 
Furthermore if $CP$ invariance holds too, then besides
$H_{11}=H_{22}$ we get also $H_{12}=H_{21}$ and consequently
$\Gamma_{12}=\Gamma_{21}={\Gamma_{21}}^*$, $M_{12}=M_{21}={M_{21}}^*$
so that $\Gamma_{ij}$, $M_{ij}$ will result all real numbers.
The time evolution of the mass eigenstates
\beq \label{eqn:eSLt}
\left( \matrix{ \vert K_S (t) \rangle \cr 
		\vert K_L (t) \rangle \cr}\right) 
=  V(t) 
\left( \matrix{ \vert K_S \rangle \cr 
		\vert K_L \rangle \cr}\right) \quad 
\eeq
is governed by the matrix elements
\beq \label{eqn:Vrs}
V_{\alpha\beta} = \langle K_\alpha \vert 
\exp \left[ -{i\over {\hbar} } \Hscr t\right] 
\vert K_\beta \rangle
\eeq
where greek letters denote $K_S$, $K_L$. The evolution matrices $U$ 
and $V$ are then related by the following similarity transformation
\beq
U=\Rscr V \Rscr^{-1}
\eeq
where the complex  scaling $\Rscr$ in any CPT invariant 
theory can be found by means of the
the eigenvalues of the effective Hamiltonian ${\cal H}$
\beq
\lambda_S = H_{11} -\sqrt {H_{12}H_{21}}=M_{11}-{i\over 2} \Gamma_{11}
-Q=m_S -{i\over 2} \Gamma_S
\eeq

\beq
\lambda_L = H_{11} +\sqrt {H_{12}H_{21}}=M_{11}-{i\over 2} \Gamma_{11}
+Q=m_L -{i\over 2} \Gamma_L 
\eeq
\cp
where
\beq
Q=\sqrt{ H_{12} H_{21} } = 
\sqrt{(M_{12}-{i\over 2} \Gamma_{12}) ({M_{12}}^*-{i\over 2} 
{\Gamma_{12}}^*)} .
\eeq
These real ($m_{S,L}$) and imaginary ($\Gamma_{S,L}$) components will define the 
masses and the decay width of the ${\cal H}$ eigenstates $K_S$ e $K_L$. 
These short- and long-lived particles result then a linear combination 
of $\ko$ and $\kbo$:
\beq 
\left( \matrix{ \vert K_S \rangle \cr 
	  \vert K_L \rangle \cr}\right) =
\Rscr^t
\left( \matrix{ \vert \ko \rangle \cr 
	  \vert \kbo \rangle \cr}\right) 
\eeq
where usually $\Rscr^t$ is parameterized according to the following 
relation
\beq
\Rscr^t =
 \left(
\matrix{
p & -q \cr
p & q \cr}
\right)
= {1\over {\sqrt{1+\vert\alpha\vert^2}}} 
\left(
\matrix{
1 & -\alpha \cr
1 &  \alpha \cr}
\right) \quad ,
\eeq
being $|p|^2 + |q|^2 = 1$. With a proper phase choice, the mixing
parameter
\beq
\alpha = {q \over p} = 
\sqrt{\frac{{H}_{21}}{{H}_{12}}} = \sqrt{
{ {M_{12}^* - {i\over 2}\Gamma_{12}^*}\over
  {M_{12} - {i\over 2}\Gamma_{12} } } }
\eeq
manifests its presence in the non orthogonality between the oblique 
independent states $K_L$ and $K_S$:
\beq
\langle K_S \vert K_L \rangle =
\frac{ {{{1 - \vert\alpha\vert^2}}} }{
{{{1 + \vert\alpha\vert^2}}} } \quad .
\eeq
The experimental evidence \cite{CCFT} in 1964 that both the short-lived 
$K_S$ and long-lived $K_L$ states decayed to $\pi \pi$ upset 
this tidy picture.
It means that the states of definite mass and lifetime are never 
more states with a definite $CP$ character.
\noindent
With the conventional choice of phase, the $CP$ eigenstates $K_1$ and 
$K_2$ enter into play so that the mass eigenstates
can be parameterized by means of an impurity complex parameter
$\epsilon$ which encodes the indirect mixing effect of CP violation 
in the neutral kaon system.
The general expression of this impurity parameter 
\beq
\epsilon={e^{i{\pi \over 4}}\over {2\sqrt {2}}} \left( {{\hbox{ImM}_{12}}\over 
{\hbox {Re}M_{12}}}\ -{i\over 2}{{\hbox {Im}\Gamma_{12}}\over 
{\hbox {Re}M_{12}}} \right)=
{e^{i{\pi \over 4}}\over {\sqrt{2} \Delta m}} \left( { \hbox {Im}M_{12} + 2 
\xi_0 \hbox{Re} M_{12} } \right) 
\eeq
where $\xi_0$ represents the additional manifestations  of 
the $CP$ violation due to the effective isospin decay amplitudes.
Neglecting these decay effects, to derive the amount of this
complex parameter, it appears more convenient to write 
\beq
\alpha = {q \over p} = {{1 - \epsilon}\over {1 + \epsilon}}
\eeq
and then
\beq \label{eqn:eps}
\eqalign{
\epsilon =& \frac{p-q}{p+q} = \frac{1-\alpha}{1+\alpha}=
\frac{\sqrt{H_{12}} - \sqrt{H_{21}}}{\sqrt{H_{12}} + 
\sqrt{H_{21}}}=\cr
&= { {2i \hbox{Im}M_{12} + \hbox{Im}\Gamma_{12}} \over 
{ (2\hbox{Re}M_{12} - i \hbox{Re}\Gamma_{12}) + 
(\Delta m -{i\over 2} \Delta\Gamma) } }
\simeq
\frac{i~{\rm Im} M_{12} + {\rm Im} (\Gamma_{12}/2)}{(\Delta m 
- {i\over 2} \Delta\Gamma)} \quad . \cr } 
\eeq
Thus it results evident that the CP-violation parameter $\epsilon$
arises from a relative imaginary part between the
off-diagonal elements $M_{12}$ and $\Gamma_{12}$ i.e. if 
$arg(M_{12}^*\Gamma_{12})\ne 0$.
Although the discovery of $CP$-violation~\cite{CCFT}
indicated that the kaon system is somewhat more complex
than the typical oscillating two-state 
problem and involves considerably more subtle complications; 
nevertheless there are many attempts to illustrate this system in
classical physics~\cite{BW}.
In this context, the origin of the space-time
discrete symmetries and their violation represent one of the
most controversial topic in the realm not only of physics. For a long time, the
fact that Maxwell  equations were invariant under space-inversion or parity
($P$) and  time-reversal ($T$) bolstered the idea that all the laws of physics 
are invariant under those discrete operations. It was easily seen that 
electromagnetic equations possess another discrete symmetry since they 
are unaffected by a charge conjugation ($C$) operation which reverses 
the sign of all the charges and converts a particle into its 
antiparticle. However, since 1964 \cite{CCFT}, we know 
that $CP$ is also violated
(although to a much lesser extent than the amount of
parity violation~\cite{Wu}) in 
weak interactions among fundamental particles.
On the 
other hand, the origin of $CP$-violation is still not explained since 
$CP$-violating tiny effects are known to be smaller than the usual 
weak interaction strength by about three orders of magnitude and it
is not excluded that $CP$-violation could be an indication of some effects 
of new physics at a much higher energy scale. The only 
almost overwhelming theoretical prejudice comes against $CPT$ 
violation.
There are very strong reasons \cite{CPT} to believe that fundamental 
interactions can never violate $CPT$ invariance. Thus, except some loosing
theoretical models, the validity of $CPT$ is assumed and consequently the $T$
violation is  supposed to be reflected immediately in a $CP$ counterpart.
However, it should be borne in mind that observation of a $T$ odd 
asymmetry or correlation is not necessarily an indication of $CP$
(or even $T$) violation. The reason for this is the anti-unitary 
nature of the time reversal operator in quantum mechanics. As a 
consequence of this, a $T$ operation not only reverses spin and 
three-momenta of all particles, but also interchanges initial and 
final states. Put differently, this means that final-state 
interactions can mimic $T$ violation, but not genuine $CP$ violation.
Presently, genuine experimental evidence of CP--violation comes from the 
mixing and decays of the neutral kaon system.
In fact, the intrinsic dissipative nature of this unstable system
and its decay sector, faced with the problem of the complex
eigenvalue of the Hamiltonian and therefore with the extension of the Hilbert
space which is a common tool to deal with open systems far from equilibrium.

In general, we can resume that 
there are three complementary ways to describe the evolution of the
complex neutral kaon system:
\medskip

{1)}{  In terms of the mass eigenstates $K_{L,S}$, which do not posses
definite strangeness
\beq
\eqalign{
\kst=&\kso\eas \qquad i\lambda_S={\it i}m_S+{{\Gamma_S}\over 2} \cr
\klt=&\klo\eal \qquad i\lambda_L={\it i}m_L+{{\Gamma_L}\over 2} \quad . \cr}
\eeq
\smallskip

{2)}{  In terms of the flavour eigenstates, whose time evolution
is more complex
\beq
\eqalign{
\kot=&\fpt\koo+{1 \over \alpha}\fmt\bkoo\cr
\bkot=&\alpha\fmt\koo+\fpt\bkoo\cr}
\eeq
with
\beq
\eqalign{
\fpmt=&{1\over 2}(\eal \pm \eas)={1\over 
2}\eal\left[1 \pm e^{-i(\lambda_S-\lambda_L)t}\right] \quad .\cr
\cr}
\eeq

{3)}{ In terms of the $CP$--eigenstates $K_1$ and $K_2$  
\beq
\eqalign{
\kku=&{1\over{\sqrt 2}}(\kk+\kbk)\cr
\kkd=&{1\over{\sqrt 2}}(\kk-\kbk)\cr}\qquad\qquad
\eqalign{
CP\kku=&+\kku\cr
CP\kkd=&-\kkd \quad ,\cr}
\eeq
which we let us express the mass eigenstates as 
\beq
\eqalign{
\kks=&{1\over{\sqrt 2}}\left[ (p+q) \kku+(p-q)\kkd)\right]\cr
\kkl=&{1\over{\sqrt 2}}\left[ (p-q) \kku+(p+q)\kkd)\right] \quad . \cr}
\eeq
\medskip
The three bases $\{K_S,K_L\}$, $\{K^0,{\overline K}^0\}$ 
and $\{ K_1$, $K_2\} $, are completely equivalent.

\section{\bf The Dissipative Effects in Coupled Oscillations.}
\bigskip

In classical physics, an analogue of the two-state mixing problem which
leads to a non-zero value of $\epsilon$ seems to strain its purport more than it
is necessary, since the equations of motion in classical mechanics are
time-reversal invariant. The main features of irreversibility enter only
considering the effects of dissipation. Anyway, these results seem to reflect
the well known requirements of additional complementarity relations,
which occur at classical level,
to make the equations of motion of dissipative systems derivable from 
a variational principle \cite{Bate}.
We can recover the theory for the non conservative systems
in classical physics
in terms of the standard formalism of analytical mechanics.
In the case of $n$
independent generalized coordinates $q_k$, the equations of
motions are given by
\beq
\frac{d}{dt} \frac{\partial {\cal T}}{\partial q_j} -\frac{\partial 
{\cal T}}{\partial t}
= {\cal Q}_j
\eeq
where the generalized momenta are of the form
\beq\label{eq:mom}
{\cal Q}_j =  \sum_{i=1}^{n}  {\cal A}_{i j} q_{i} + {\cal B}_{i j} 
\frac{dq_i}{dt}\; .
\eeq
In the special case of conservative system, a configuration of stable
equilibrium is specified by the following Lagrangian equations
\beq
\frac{d}{dt} \frac{\partial {\cal T}}{\partial {\dot q}_k} +\frac{\partial
{\cal U}}{\partial q_k} = 0 \quad \mbox{or} \quad \frac{\partial {\cal 
L}}{\partial q_k} -
\frac{d}{dt} \frac{\partial {\cal L}}{\partial {\dot q}_k}  = 0 \quad \;  
{\cal L}={\cal T}-{\cal U} \quad .
\eeq
As usual, the kinetic tensor
\beq
{\cal T}= \frac{1}{2} \sum_{j,k} {\cal T}_{jk} {\dot q}_j {\dot q}_k
\eeq
and the potential function
\beq
{\cal U}= \frac{1}{2} \sum_{j,k} {\cal U}_{jk} q_j q_k 
\eeq
are supposed bilinear without any loss of generality.
Of course, the exact form of (${\cal T}_{ij}$) and (${\cal U}_{ij}$) specifies
the way in which
the motions of the various coordinates are coupled. The problem of determining
the normal solutions reduces to that of simultaneous diagonalization of these 
two real symmetric matrices associated with each of the kinetic and potential
energies. The problem consists practically in finding a coordinate
transformation which simultaneously diagonalizes both (${\cal T}_{ij}$) and 
(${\cal U}_{ij}$).
This is equivalent to the following set of second order homogeneous
differential equations
\beq\label{eq:mot}
\sum_j ({\cal U}_{jk} q_j + {\cal T}_{jk} {\ddot q}_j ) = 0 \quad .
\eeq
This underlying mathematical problem can be formulated elegantly in matrix
language since it can be compactly associated with the following Lagrangian
\beq
{\cal L}={\cal T}-{\cal U}= \frac{1}{2} \left[ {\dot {\vec q}}^t 
{\cal T} {\dot {\vec q}} -
{\vec q}^t {\cal U} {\vec q} \right] \quad .
\eeq
Then the problem to determine the normal modes reduces to an eigenvalue problem
involving the simultaneous diagonalization of two real symmetric matrices,
classically associated with the kinetic and potential energies. The recipe
for carrying out this diagonalization is straightforward. The kinetic and
potential energy matrices are simultaneously diagonalized by means of a sequence
of three transformations from the original to the normal coordinates 
${\bxi}$
\beq 
{\vec q}\ =\  R_1\  {\vec q^\prime} \ = \ R_1\  S\  {\vec q^{\prime\prime}}\ =\
R_1 \ S \ R_2\ {\bxi} \quad .
\eeq
In this new normal coordinates ${\bxi}$, the equations of motion are
uncoupled. The net transformation matrix $G\ =\ R_1\ S \ R_2$ is not orthogonal
because $S$ is not orthogonal. This matrix $G$ represents the most general
(homogeneous) linear transformation which includes two rotation matrices $R_1$,
$R_2$ and a contraction $S$.
We have to sacrifice the property of orthogonality in order to construct a
matrix that can simultaneously diagonalize both ${\cal T}$ and ${\cal 
U}$. The effect of
this transformation is to diagonalize the kinetic and potential energy 
matrices:
\beq\eqalign{
\hbox{diag}\ {\cal T}\ =&\ G^t\ {\cal T}\ G \cr
\hbox{diag}\ {\cal U}\ =&\ G^t\ {\cal U}\ G \quad .\cr }
\eeq
To be more explicit, in the case of normal solution of the form
$q_j = a_j e^{i (\omega t - \delta)}$
we obtain $n$ linear homogeneous algebraic equations:
\beq
\sum_j ({\cal U}_{ij} q_j - \omega^2 {\cal T}_{ij}) a_j = 0 \quad ,
\eeq
where the common phase factor $e^{i (\omega t - \delta)}$ has been cancelled. A
non trivial solution corresponds to the secular equation
\beq
\det ({\cal U}_{ij} - \omega^2 {\cal T}_{ij} ) = 0 \quad .
\eeq
In general, this n-order spectral equation in the eigenfrequencies $\omega^2$
cannot be solved simply.
Nevertheless, in the case of dissipative systems the coupled differential
equations that describe the damped motion can be obtained only by means of the
use of the generalized momenta in Eq.\ (\ref{eq:mom}).
In general we must consider three matrices ${\cal T}_{ij}$, 
${\cal A}_{ij}$ and ${\cal B}_{ij}$.
A normal mode solution of the kind $q_j = {\cal Q}_j e^{-i \omega t}$
can be derived from the secular equation
\beq
\det ( T_{ij} \omega^2 + A_{ij} + i \omega B_{ij})=0 \quad .
\eeq
Using the modern Dirac notation, although we are discussing classical
quantities, its physical solution derives from the characteristic equation
\beq
{\cal H} {| \bf q >} = \omega^2 {| \bf q >}~~~,
\eeq
being
\beq
{\cal H} = i \omega {\cal B} - {\cal A}~~~.
\eeq
Thus the equations of the motion can be rewritten as
\beq
{\cal T} \frac{d^2 | {\bf q} >}{dt^2} = {\cal A} | {\bf q} > 
- i{\cal B} \frac{d| {\bf q} >}{dt} \, .
\eeq
This set of equations resembles the case of a system of coupled damped harmonic
oscillators \cite{Vitiello}. The dissipative effects are parameterized
by the presence of small off-diagonal terms in ${\cal H}$,
as it was carefully illustrated in a recent publication \cite{LT} in
the classical framework of mechanics and by means of electromagnetic
coupled circuits. The time-reversed not-invariant processes can be
induced by dissipative  effects which yield a not vanishing imaginary
part for the relevant Hamiltonian. 

\section{\bf Electromagnetic Constitutive Relations.}

The description of the
fundamental nature of the electromagnetic interactions induced in
a medium is obscured by several mathematical problems \cite{OSR}.
These difficulties are simply an expression of our ignorance about
the dynamics of all processes going on in the medium or our inability
to quantify them in detail. From a purely macroscopic point of view,
the crucial question is to append certain {\it constitutive relations}
in dependence with an effective parameterization of the induced charges
and current densities and feeding them into the Maxwell's equations.
The presence of the induced sources is taken into
account by introducing a few empirical constants in the equations.
The number of such constants necessary depends on the nature of the
medium. For the simpler linear isotropic materials, it is well known that
one needs two scalar quantities: the dielectric constant
$\varepsilon$ and the magnetic permeability $\mu$, to describe all
phenomena of macroscopic electrodynamics, according to the relations
		    \begin{eqnarray}\label{eqn:epsmu}
\vec D \equiv \varepsilon_0 \vec E + \vec P = \varepsilon \vec E \quad,\quad
\vec H =  \frac{\vec B}{\mu_0} - \vec M = \mu^{-1} \vec B \,.
		    \end{eqnarray}
Once the empirical constants $\varepsilon$ and $\mu$
are known, one can then derive the electric $\vec E$ and magnetic 
$\vec B$ vector fields.
However, for complex systems one needs in general more than these
two constants to describe electrodynamical phenomena in them. 
In many instances this inadequacy may be
addressed by postulating a more general
set of constitutive relations to replace those in
Eq.\ (\ref {eqn:epsmu}):
		\begin{eqnarray}
\vec D &=& \varepsilon \vec E + \beta \vec B   \,, \label{D=EB}\\
\vec H &=& \mu^{-1} \vec B + \gamma \vec E  \,. \label{H=EB}
		\end{eqnarray}
We should comment that the restriction of the number of 
electromagnetic constants depends on the nature of the medium 
in question, since in the most general case they are rank-2 tensors 
\cite{OSR} .
While the approach of postulating these general constitutive
relations are well known~\cite{OSR}, until now, it is not deeply discussed
in connection with the symmetry requirements which could be of great help
in analyzing the dynamics of propagation.
The symmetry requirements under spatial translation or under
rotations are well-documented experimentally, and no physicist
has serious doubts about their validity as a consequence of the
conservation of momentum and angular momentum. Invariance under spatial
inversion and time reflection, on the other hand, seems to hold only
approximately, for some classes of phenomena.
The point is that spatial inversion symmetry, which is associated
with parity ($P$) invariance, is preserved by strong and
electromagnetic interactions but violated by the weak force~\cite{Wu}.
Since the latter interaction is at least
six orders of magnitude weaker than its strong and electromagnetic
counterpart, parity violation can generally be neglected.
Similar comments apply to the case
of time reversal ($T$) symmetry, which is found~\cite{CCFT} to be
violated  only by an effect even more feeble than the weak force.
Maxwell's equations have
another discrete symmetry since they are unaffected by the
operation under which
$\rho$, ${\vec j}$, ${\vec E}$ and ${\vec B}$ all change sign.
This operation,
called charge conjugation ($C$) because the sign of all charges
are reversed, plays an important role in quantum field theory,
where it converts a particle into its associated antiparticle.
The transformation properties of the electromagnetic fields,
potentials and currents, with respect to the discrete symmetries
$C$, $P$ and $T$, and some important combinations like $CP$ and
$CPT$ (Table \ref{t:CPT}), can be better
understood to consider the Fourier-space version of the Maxwell's
equations.

\noindent
In fact in many field of physics, including electrodynamics, it is
convenient to re-write the (Maxwell) differential equations of
motion in terms of their Fourier components.
In this linear approximation, from the usual results of
Fourier analysis, we express a general solution
as a superposition of plane waves through the following relation
		\begin{eqnarray}
{\vec E}(t, \vec x) = {1 \over (2\pi)^2} \int d\omega \int d^3 \vec k \,
\widetilde{\vec E} (\omega, \vec k) \exp [i(\vec k \cdot \vec x -
\omega t)]    \label{fourier}
		\end{eqnarray}
and similarly for other quantities. Now, if the Fourier
transform of a linear equation is taken, the space and time
drop out. By these means, the set of coupled (Maxwell)
differential equations are replaced by a set of coupled
algebraic equations, which of course greatly reduces the
difficulty of the problem. The solutions of the algebraic
equations are the Fourier transforms of the required
quantities which can then be obtained by Fourier inversion.
The Fourier momentum space version of the Maxwell's equations in a
medium is then given by
		\begin{eqnarray}
\vec k \cdot \widetilde{\vec B} =0 \quad &,&\quad  \vec k \times 
\widetilde{\vec E} =
\omega \widetilde{\vec B}\label{MaxEq:hom}  \\
i\vec k \cdot \widetilde{\vec D} = \widetilde{\rho}\ext \quad&,&
\quad i \vec k
\times \widetilde{\vec B} = \mu_0 \widetilde{\vec J} -
i\frac{\omega}{c^2} \widetilde{\vec E} 
\label{MaxEq:DH}
		    \end{eqnarray}
\noindent
with source terms just the external charges and currents. In writing 
these relations it is assumed that the response of the system to the 
applied electromagnetic fields is linear. Even when non linear effects 
are considered, they could be usually treated as corrections to linear 
ones. In this approach, it is assumed that all quantities which vary 
with time and space differ only slightly from an average value.
The source Maxwell's equations 
\ (\ref{MaxEq:DH}) reduce to the wave form

\beq
\frac{\omega^2}{c^2} \widetilde{\vec E}+ \vec k \times \left( \vec k 
\times \widetilde{\vec E} \right) = -i \omega \mu_0 \widetilde{\vec J}
\quad .
\eeq

\begin{table} \caption[]
{{\small\sf
The transformation properties of the electromagnetic currents,
fields and potentials, under the discrete symmetries $P$, $T$
and $C$, and some important combinations like $CP$ and $CPT$.
If for a generic ${\tilde \Phi}$ of these quantities, including $\omega$
and $\vec k$ themselves, we write the above transformation rules
in the form
${\tilde \Phi} \stackrel{X}{\longrightarrow} \eta_X {\tilde \Phi} \, , $
then the value of the multiplicative constants $\eta_X$ is given for the
various transformations.
The same procedure is followed for the pulsation and wave vector that
transforms, for example, under the time reversal symmetry, according to
$
(\omega, \vec k) \stackrel{T}{\longrightarrow} (\omega, \vec k)\,.
$ }
\label{t:CPT}} 

\begin{center} 
\begin{tabular}{|c||c|c|c|c|c|} \hline &
\multicolumn{5}{c|}{Transformation} \\ 
\hline Quantity & $\eta_P$ & $\eta_T$ & $\eta_C$ & $\eta_{CP}$ & $\eta_{CPT}$ \\
\hline $\omega$ & $+$ & $-$ & $+$ & $+$ & $-$ \\ 
       $\vec k$ & $-$ & $+$ & $+$ & $-$ & $-$ \\ 
       $\rho$ & $+$ & $+$ & $-$ & $-$ & $-$ \\ 
       $\vec j$ & $-$ & $-$ & $-$ & $+$ & $-$ \\ 
       $\vec E$ & $-$ & $+$ & $-$ & $+$ & $+$ \\ 
       $\vec B$ & $+$ & $-$ & $-$ & $-$ & $+$ \\ 
       $\vec A$ & $-$ & $-$ & $-$ & $+$ & $-$ \\ \hline
  \end{tabular} \end{center}
\end{table}

\noindent
Eliminating the induced part ${\widetilde{\vec J}_{ind}}$ of the current 
${\widetilde{\vec J}}={\widetilde{\vec J}_{ind}}+{\widetilde{\vec J}_{ext}}$
in terms of general conductivity tensor ${\widetilde {\sigma_{ij}}}$

\beq
\left({\widetilde{\vec J}_{ind}} \right)_i ={\widetilde {\sigma_{ij}}}
{\widetilde{\vec E}}_j
\eeq
\noindent
then, Eq.(51) reduces to the inhomogeneous wave equation

\beq
\widetilde{\Lambda}_{ij} {\widetilde{\vec E}}_j =
-\frac{\mu_0 c^2}{\omega} \left( {\widetilde {\vec J}}_{ext} \right)_i,
\eeq
\noindent
with

\beq
\widetilde{\Lambda}_{ij} \left( \omega , \vec k \right) = 
\frac{k^2 c^2}{\omega^2} \left( k_i k_j 
-\delta_{ij} \right) + {\widetilde K}_{ij} \left( \omega , \vec k \right)
\eeq
\noindent
being the equivalent response tensor

\beq
{\widetilde K}_{ij} \left( \omega , \vec k \right) = 
\delta_{ij} + \frac{i}{\omega \epsilon_0} 
{\widetilde {\sigma_{ij}}}\left( \omega , \vec k \right) \quad .
\eeq
\noindent
If the medium is dissipative, the tensor ${\widetilde K}_{ij}$ 
may be separated into a hermitian and an antihermitian part

\beq
{\widetilde K}_{ij} ={\widetilde  K}_{ij}^H +{\widetilde K}_{ij}^A
\eeq
\noindent
with
\beq
{\widetilde K}_{ij}^H = \frac{1}{2} 
\left( {\widetilde K}_{ij} + {\widetilde K}_{ij}^{\ast} \right)
\eeq

\beq
{\widetilde K}_{ij}^A = \frac{1}{2} 
\left({\widetilde K}_{ij} - {\widetilde K}_{ij}^{\ast} \right)
\eeq
\noindent
which corresponds to a separation into a dissipative and
non-dissipative part.
\noindent
The Kramers-Kronig relations assure that the hermitian and antihermitian
parts are not independent. In fact,

\beq
{\widetilde K}_{ij}^A = \frac{i}{\pi} {\cal P} \int_{- \infty}^{+ \infty}
\frac{{\widetilde K}_{ij}^H- \delta_{ij}}{\omega - \omega'} \, d \omega'
\eeq

\beq
{\widetilde K}_{ij}^H - \delta_{ij} = \frac{i}{\pi} {\cal P} \int_{- \infty}^{+ \infty}
\frac{{\widetilde K}_{ij}^A}{\omega - \omega'} \, d \omega' \quad .
\eeq
\noindent
For an isotropic medium, that is only spatially dispersive, we have three 
independent (transverse, longitudinal and rotatory) parts in $K_{ij}^H$

\beq
{\widetilde K}_{ij}^H = \varepsilon^T \left( \delta_{ij} - k_i k_j \right) + 
\varepsilon^L k_i k_j + \varepsilon^R \epsilon_{ijr}k_r = 
\left(
\matrix{
\varepsilon^T & \varepsilon^R & 0 \cr
-\varepsilon^R & \varepsilon^T & 0 \cr
0 & 0 & \varepsilon^L \cr}
\right)
\eeq
\noindent
where the last matrix form holds if we choose $\vec k$ along the $z$-axis.
\noindent
The fields $\widetilde{\vec E}$ and $\widetilde{\vec B}$ depend on 
$(\omega, {\vec k})$, whereas the quantities $\varepsilon^T$,
$\varepsilon^L$ and $\varepsilon^R$, which represent the properties
of the medium, are supposed to depend on 
$\vec k$ only through its magnitude $k
\equiv\left|\vec k\right|$. In fact we are considering a medium in
which only the amplitudes of the electric and magnetic polarizations
depend on direction. The electromagnetic constants $\varepsilon^T$,
$\varepsilon^L$ and $\varepsilon^R$
have no anisotropy and are the same in all the directions.
Before we proceed
any further, we want to note that there is one relation that
$\varepsilon^T$, $\varepsilon^L$ and $\varepsilon^R$ must satisfy
based on very general grounds. It is simply the
requirement that the  fields $\vec E$, $\vec B$ and the current $\vec j$ are, in
coordinate space, real quantities. This yields that, for real values of
$\omega$, the electromagnetic constants satisfy

		\begin{eqnarray}
\label{hermiticity}
f^\ast(-\omega, {\vec k}) = f(\omega, {\vec k}) \,,
		\end{eqnarray}
where $f$ stands for any of the electromagnetic constants $\varepsilon^T$,
$\varepsilon^L$ and $\varepsilon^R$ and the dependence is
supposed through its magnitude $k$, as noted earlier. All that means that, the 
real parts of these electromagnetic constants are even in $\omega$ 
whereas the imaginary parts are odd.
We stress that this is a very general property, not tied up in any way 
with the discrete $C$, $P$ and $T$ symmetries of space-time.
Of course, for the validity of causality and the consequent 
Kramers-Kronig dispersion relations, $\omega$ can in general be allowed 
to be complex and this feature must be included in the analysis.

\noindent
The implication of $C$, $P$ and $T$ symmetries on the electromagnetic 
constants $f$ imposes that the reversed equations result invariant under the   
$C$, $P$ and $T$ constraints if
		\begin{eqnarray}
\mbox{$C$ invariance} \Rightarrow \mbox{no constraints on $\varepsilon^T$, $\varepsilon^L$, and
$\varepsilon^R$} \, ,
		    \end{eqnarray}

		\begin{eqnarray}
\mbox{$P$ invariance} \Rightarrow \varepsilon^R = 0 \, ,
		    \end{eqnarray}

		\begin{eqnarray}
\mbox{$T$ invariance} \Rightarrow  \left\{ \begin{array}{rcl}
\varepsilon^T \left( -\omega, {\vec k} \right) &=& \varepsilon^T \left
( \omega, {\vec k} \right)\\
\varepsilon^L \left( -\omega, {\vec k} \right) &=& \varepsilon^L \left
( \omega, {\vec k}
\right)\\
\varepsilon^R \left( -\omega, {\vec k} \right) &=& - \varepsilon^R \left
( \omega, {\vec k}
\right)
\end{array} \right. \quad ,
\label{Trules}
	   \end{eqnarray}
\noindent
and being in general $f^\ast (-\omega, {\vec k})=f(\omega, {\vec k})$, 
thus implies 
that $\varepsilon^R$ must be purely 
imaginary and an odd function of $\omega$, whereas $\varepsilon^T$ and 
$\varepsilon^L$ should be real and even function of $\omega$.
In the case of a dissipative media, the response tensor components will
respect the Onsanger relations

\beq
\eqalign{
K_{ij}^H \left(-\omega, {\vec k} \right) =& K_{ij}^H \left( \omega, {\vec k} 
\right) \cr
K_{ij}^A \left(-\omega, {\vec k} \right) =& -K_{ij}^A \left( \omega, {\vec k} 
\right) \cr 
}
\eeq
\noindent
whereas in general $K_{ij}\left( \omega, -{\vec k} \right) = K_{ij}\left( 
\omega,
{\vec k} \right)$.

\noindent
$CP$ invariance obviously gives the same constraint as parity alone
does, since charge conjugation does not affect.

		\begin{eqnarray}
\mbox{CPT invariance} \Rightarrow
\left\{   \begin{array}{rcl}
\varepsilon^T \left( - \omega, {\vec k} \right)&=&\varepsilon^T \left(
\omega, {\vec k} \right)\\
\varepsilon^L \left( -\omega,{\vec k} \right)&=&\varepsilon^L \left( \omega, 
{\vec k}
\right)\\
\varepsilon^R \left( -\omega,{\vec k} \right)&=& \varepsilon^R \left( \omega, 
{\vec k}
\right)
\end{array}  \right. \quad .
	   \end{eqnarray}
\noindent
From these relations, it is quite clear that the presence of 
the $\varepsilon^R$-term implies some properties that are asymmetric under  
$P$ and $CP$ transformations. Similarly, since $T$ and $CPT$ imply mutually 
inconsistent constraints on $\varepsilon^R$, at least one of them must be 
violated in order for $\varepsilon^R$ to exist.
$\varepsilon^R$ might be expected to vanish if all interactions conserve $C$, 
$P$ and $T$.
However since 1957~\cite{Wu} we know that parity is violated by weak 
interactions and since 1964~\cite{CCFT} we know that $CP$ is also violated
although to a much lesser extent. On the other hand
the induced currents arise because of complicated processes 
taking place within the medium, including (super) weak interactions, so that
there is no reason why the $\varepsilon^R$-term should not be present. Of 
course since the $\varepsilon^R$-term (if present) violates $CP$ it would seem to 
be extremely tiny. The only constraint on the value of $\varepsilon^R$ seems 
$\varepsilon^R(\omega, k)=\varepsilon^R(-\omega, k)$ which follows from $CPT$ symmetry, 
since there are strong reasons to believe that fundamental 
interactions between particles can never violate $CPT$.
But in the presence of matter, the $CPT$ breaking effects could be relevant 
due to the effects of the medium in the underlying effective theory.
So that, $CPT$ asymmetries could be large and their invariant constraints 
become irrelevant.
As a result, parity violation in the weak 
interactions of the medium can give rise to a non vanishing value of $\varepsilon^R$
and similarly a much smaller $T$-violating 
effects should be taken also seriously into account. As we found 
before in Eq.\ (\ref{Trules}), if  $T$-invariance holds, $\varepsilon^R$ must be purely 
imaginary and an odd function of $\omega$.
Then, it is worth noting to the extent that $T$-violation can be 
neglected in the fundamental interactions, it is conceivable to think 
a medium where $\varepsilon^T$ and $\varepsilon^L$ are constants, (being both even 
functions of $\omega$), but $\varepsilon^R$ cannot have a non zero constant 
term.
In this case a medium becomes dispersive according only to $\varepsilon^R$. 
When we consider more complicated media, it is possible to obtain  
$P$-asymmetries from the medium itself. Therefore it is important to 
consider the passage of electromagnetic radiation through a generic 
medium with a built-in asymmetry. This is the case, for example, of 
natural sugar solution, where one helicity of the radiation is in 
excess over the other, and physically, it manifests itself as the 
phenomenon of optical activity.

\section{\bf The Optical Analogue of the $\kok$ Complex System.}

We seek in electromagnetism an analogue of the two-state mixing
problem which leads to a non-zero value of an optical parameter which could
be associated to the impurity parameter $\epsilon$.  The problem is quite
difficult since the Maxwell's equations are time-reversal invariant.
The main features of irreversibility enter only considering the effects of
anisotropy and dissipation. 
In an isotropic medium, transverse electromagnetic waves, have two degenerate
states of polarization, and in anisotropic media, the natural modes
are non degenerate and not necessarily transverse.
Radiation propagating through such a medium splits into components in the
two natural modes and the difference between the complex refractive indices 
of the two modes produces radiation with interfering effects when the components are recombined.
In the case of optically active media, the dynamic behavior is connected
to dissipation more than anisotropy, as far as their electromagnetic properties are concerned.
\noindent
We review here the fundamental mathematical description of the 
propagation of optical polarization in a medium.
In a complex dielectric material, the nature of the wave propagation 
is learned, in general, from the dispersion relations. These are the relations
that $\omega$ and ${\bf k}$ must satisfy in order to solve the 
homogeneous wave equation for the ${\bf E}$ and ${\bf B}$ fields: 

\beq
{\widetilde \Lambda}_{ij} {\widetilde E}_j = 0 \quad .
\eeq
\noindent
In a general complex medium, the existence of a solution is connected to the 
dispersion relation

\beq
\hbox{det} {\widetilde\Lambda}_{ij}=0 \quad .
\eeq
\noindent
When this dispersion relation is satisfied, there exists a solution of the
homogeneous wave equation.
In the vacuum, this vector solution $\FEbo$ is perpendicular to the direction of propagation of the
electromagnetic wave. 
Thus, one can discuss this solution (with complex coefficients) 

\beq
\FEbo = {\widetilde E}_1 \epsx + {\widetilde E}_2 \epsy
\eeq
in terms of the components
\beq
\eqalign{
\widetilde{E_1} = \widetilde{E_x} =& \Re \ \left( 
\widetilde{\Ebo} \cdot \epsx \right)  \cr
\widetilde{E_2} = \widetilde{E_y} =& \Re \ \left( 
\widetilde{\Ebo} \cdot \epsy  \right) \cr}
\eeq

\noindent
of the electric field transverse to the coherent beam propagating in the 
$\hat {\bf{k}} = \hat {\bf{z}}$ direction.
In this description of polarization, we refer to two basic polarization
orthonormal unit vectors $\epsx$ and $\epsy$, both being perpendicular to the wave
unit vector $\hat {\bf{k}}$.
An equivalent set of basic vectors is given by the following right and the left 
circular basis

\beq
\eqalign{
\epsr =& \frac {1}{\sqrt 2} \left( \epsx  + i \epsy  \right) \cr
\epsl =& \frac {1}{\sqrt 2} \left( \epsx  - i \epsy  \right) \quad . \cr}
\eeq
Since $\epsy=\hat {\bf{k}}\times \epsx$, these vectors are of the form
$\epsx\pm {\it i}\hat {\bf{k}}\times \epsx$. 
In this language, the effect of parity
transformation is to reverse the sign of the transverse polarization vector
$\epsy$, in a suitable gauge. Whereas, under parity, the circular polarization
vectors interchange: $\epsr\leftrightarrow \epsl$.
In non dispersive media ($\varepsilon^R=0$),
the two transverse modes have the same
physical characteristics, with the same speed of propagation
$c/\sqrt{\epsilon\mu}$ and, if parity is a good symmetry, there would be also no
difference in the physical properties of a right-circular and left-circular
polarization state of the transverse electromagnetic wave.
Wherever $\varepsilon^R = 0$, in an isotropic homogeneous medium, the
response tensor is given by

\beq
{\widetilde K}_{ij} = \varepsilon^L {\hat {\vec k}}_i {\hat {\vec k}}_j +
\varepsilon^T \left( \delta_{ij} - 
{\hat {\vec k}}_i {\hat {\vec k}}_j \right) \quad .
\eeq
\noindent
In this case, the dispersion relation 

\beq
\mbox{det} {\widetilde \Lambda}_{ij}=
\mbox{det} \left[\varepsilon^L {\hat {\vec k}}_i {\hat {\vec k}}_j +
\left( \varepsilon^T - n^2 \right) 
\left( {\hat {\vec k}}_i {\hat {\vec k}}_j -\delta_{ij} \right) \right]=
\varepsilon^L \left( \varepsilon^T - n^2 \right)^2 =0
\eeq
\noindent
yields a longitudinal dispersion relation

\beq
\varepsilon^L \left( \omega, k \right) =0 
\eeq

\noindent
and a transverse dispersive equation

\beq
n^2 = \varepsilon^T \left( \omega, k \right) = 
\frac{\epsilon \left( \omega \right)
\mu \left( \omega \right)}{\epsilon_0 \mu_0} = c^2 \epsilon \left( \omega \right)
\mu \left( \omega \right) \quad .
\eeq
\noindent
It is worth noting that only in an {\it isotropic} medium, all solutions
are polarized strictly transverse.
However, in an isotropic optically active media
$(\varepsilon^R \ne 0)$, parity, and, in a less extent time-reversal
are broken. If also dissipative effects are considered,
the two transverse, (and circular), birefringent (or dichroic) modes 
have different propagating properties. For
example, they travel with different speeds through matter and have different
indices of refraction and coefficients of absorption.
It is evident that the two transverse modes have different properties
when $\varepsilon^ R$ does not vanish.
The effect of dissipation in an isotropic medium will cause a non trivial 
complex form of $\varepsilon^ R \left( \omega, k \right)$.
In fact,
if time-reversed invariance is satisfied, $\varepsilon^T$ and $\varepsilon^L$ are real, whereas
$\varepsilon^R$ is purely imaginary only neglecting dissipation. 
In this case, for a large range of $\varepsilon^R$, the
dispersion relation is solved for a real $\omega$ and
$k$. Thus, the amplitude of the wave will remain unchanged during the
propagation, which means that there is no absorption in the medium.
In the more general case of dissipation, time-reversal symmetry can be violated and
the space-time evolution of the transverse electric field becomes governed by a
set of coupled second order differential equations

\beq
{\widetilde {\vec E}}_i = {\widetilde \Lambda}_{ij} {\widetilde {\vec E}}_j  \quad .
\eeq
\noindent
The two transverse modes no longer have the same dispersion relation. 
The transverse components are governed by the following inverse matrix
propagator

\beq
{\widetilde \Lambda}_{ij}^{\bot}= - \frac{k^2 c^2}{\omega^2}\delta_{ij} + 
\left( \delta_{ij} + \chi_{ij} \right)
\eeq
\noindent
which is defined for the transverse ${\widetilde {\vec E}}$ components
Eq. (70) with respect to ${\hat {\vec k}}$.
However, the degree and type of the polarization of 
this resulting transverse electric wave
is associated with a unit vector and therefore with a point on the unit sphere
\cite{Stokes}. The stereographic projection of this unit vector onto the
equatorial plane can be consequently associated to the ratio of the complex
components of a Poincar\`e spinor \cite{Poincare}. Nothing else that the
usual isomorphism between the rotation group and its covering group $SU(2)$. The two-components
Poincar\`e spinor requires the standard matrix representation in terms of the
Pauli matrices and it is associated to a vector representing
the angular momentum for a spin-$\frac{1}{2}$ state. In fact, this
vector results just the generator of the internal rotations and
it is called Jones vector \cite{Jones}, when it is considered in the Cartesian
basis. Its components give the Cartesian components of the complex polarized
electric field. This $2\times 2$ unimodular matrix representation of the Pauli
algebra to describe the polarization, can be replaced by the $4\times 4$ Mueller
matrix method \cite{Mueller}. Another modern method using $2\times 2$ matrices to
represent polarization is represented by Fano's density operator approach
\cite{Fano} analogously to the case of spin density theory for the
spin-$\frac{1}{2}$ particles.
It is worth noting that while the fundamental mathematical description of spin
$\frac{1}{2}$ and optical polarization are practically the same, their physical 
interpretations are quite different. In the description of spin $\frac{1}{2}$
we have a spin vector whose direction in a cartesian 3-dimensional space depends
upon a state vector in a complex 2-dimensional (spin-up, spin-down) spinor
space. Anyway, in the description of polarization the two ordinary components $x$
and $y$ play the same role which the spinor bases played while describing
spin $\frac{1}{2}$.
Although the problem to describe the evolution of the two transverse polarization
states of the electromagnetic radiation propagating into a complex medium
can be described in these several manners, nevertheless, a dispersion relation
governing the propagation is required.

\noindent
In our case of a wave propagating inside 
a medium with complex $\varepsilon^R \neq 0$, the space propagation of 
the transverse optical polarization states is governed by 
the transverse response tensor

\beq
K_{ij}^{\bot} = \delta_{ij} + \chi_{ij} = {\cal D}^2
\eeq
\noindent
which is given in terms of a general non Hermitian susceptibility tensor
\beq
\chi =
\left(
\matrix {\chi_{11} & \chi_{12} \cr
\chi_{21} & \chi_{11} \cr}
\right) \quad .
\eeq
\noindent
Consider a plane polarized wave incident on a medium at $z=0$, moving in the
positive $z$-direction, the transverse electric components can be assumed in a normal form of
the type $ \Ebo (z,t) = \Ebo (z) e^{-i \omega t}$ where
\beq
\Ebo (z) = E_1 (z) \epsx + E_2 (z) \epsy
\eeq
\noindent
\beq
\left(
\matrix {E_1 (z) \cr
E_2 (z) \cr}
\right)=
\exp \left[ - i z {\cal D} \right] 
\left(
\matrix {E_1 (0) \cr
E_2 (0) \cr}
\right)
\eeq
\noindent
where
\beq
{\cal D} =  \frac{\omega\sqrt{\mu \epsilon}}{c}
\left(
\matrix{1+ \chi_{11} & \chi_{12} \cr
	 \chi_{21} & 1+ \chi_{11} \cr}
\right)^{\frac{1}{2}} =\frac{\omega\sqrt{\mu \epsilon}}
{ c ( \sqrt{\lambda_1} + \sqrt{\lambda_2} ) }
\left(
\matrix{1+ \chi_{11} + \sqrt{\lambda_1 \lambda_2} & \chi_{12} \cr
	 \chi_{21} & 1+ \chi_{11} + \sqrt{\lambda_1 \lambda_2} \cr}
\right)
\eeq
\noindent
In the matrix square-root we use the eigenvalues of ${\cal D}^2$
\beq
\lambda_{1,2} = \frac{1}{2} \left( \Sigma \pm \Delta \right) 
\eeq
\noindent
with
\beq
\eqalign{
\Sigma=& \mbox{tr}{\cal D}^2 = 
2 + \mbox{tr} \chi = 2 \left( 1 + \chi_{11} \right) \cr
\Delta^2 = & \mbox{tr}^2 {\cal D}^2 - 4 \mbox{det}{\cal D}^2 = 
\mbox{tr}^2 \chi - 4 \mbox{det} \chi =
4 \chi_{12} \chi_{21} \quad . \cr
}
\eeq
\noindent
Therefore, the electric field will propagate as
\beq
\eqalign{
{E}_+ (z) = &{E}_+ (0) e^{- {\it i} \lambda_+ z} \cr
{E}_- (z) = &{E}_- (0) e^{- {\it i} \lambda_- z} \cr }
\eeq
\noindent
where the $\cal D$ eigenvalues $\lambda_+$ and $\lambda_-$ are given by
\beq
\lambda_{\pm}=\lambda_{1,2} + \sqrt{\lambda_1 \lambda_2} \quad .
\eeq
\noindent
Here, ${E}_+ $ and ${E}_-$ are given by
\beq
\eqalign{
{E}_+  =& 
{1\over {\sqrt{1+\vert\alpha\vert^2}}} 
\left ( {E}_1   
+  \alpha {E}_2  \right ) \cr
{ E}_- =& 
{1\over {\sqrt{1+\vert\alpha\vert^2}}} 
\left ( {E}_1 
-  \alpha {E}_2 \right ) \quad . \cr}
\eeq
\noindent
We recognize that the parameter $\alpha$ could be put in relation
with the off-diagonal terms of the propagation matrix $\cal D$ by the
following equation
\beq
\alpha =\sqrt{\frac{ D_{21}}{D_{12}}}=
\sqrt{\frac{K_{21}^{\bot}}{K_{12}^{\bot}}}
=\sqrt{\frac{K_{12}^{H \ast} - K_{12}^{A \ast}}
{K_{12}^H + K_{12}^A}}
=\sqrt{\frac{\chi_{21}}{\chi_{12}}} \quad .
\eeq
\noindent
It can be viewed that it depends upon the off-diagonal terms of the
susceptibility tensor $\chi$.
In a non-dissipative medium $K_{ij}^A = 0$ and $\vert\alpha\vert=1$.
In particular, in an isotropic medium which is only dispersive, we 
have: $\alpha = i$.
The analogy with the neutral kaon system is quite evident;
the eigenstates the space propagation
resemble the mass eigenstates of the effective Hamiltonian $\cal H$
governing the evolution of $\kl$ and $\ks$.

\noindent
In summary, it was presented a successful attempt to realize the 
optical analogue which reproduces the peculiar asymmetric effects of 
the $CP$--violating propagation. The absorptive and dispersive 
contributions in isotropic media can then resemble the intriguing 
features of the kaon complex evolution in the realm of high energy 
particle physics. A last remark merits the eventuality to introduce an 
absorptive effect by means of the space
anisotropy whose consequent longitudinal propagation, however,
seems outside an abelian and strictly massless effective theory
of the electromagnetic interactions.

\end{document}